\documentclass[twocolumn,twoside,slac_two,revsymb]{revtex4}
\usepackage{graphicx}
\usepackage{amssymb}
\usepackage{fancyhdr}
\begin{document}

\title{Identifying the neutrino mass hierarchy with supernova neutrinos}

%

\author{R. Tom\`as}
\affiliation{AHEP Group - Institut de F\'{\i}sica Corpuscular (CSIC -
  Universitat de Val\`encia)\\ 
Apartat de Correus 22085, E-46071 Val\`encia, SPAIN}
%

\begin{abstract}
We review how a high-statistics observation of the neutrino signal
from a future galactic core-collapse supernova (SN) may be used to
discriminate between different 
neutrino mixing scenarios. Most SN neutrinos are emitted in the
accretion and cooling phase, during which the flavor-dependent
differences of the emitted neutrino spectra are small and rather
uncertain. Therefore the discrimination between neutrino mixing
scenarios using these neutrinos
should rely  on observables independent of the
SN neutrino spectra. We discuss two complementary methods that allow
for the positive identification of the mass hierarchy without knowledge of 
the emitted neutrino fluxes, provided that the 13-mixing angle is
large, $\sin^2\theta_{13}\gg 10^{-5}$. 
These two approaches are the observation of modulations in the
neutrino spectra by Earth matter effects or by the passage of
shock waves through the SN envelope.
If the value of the 13-mixing angle is unknown, using additionally the
information encoded in the prompt neutronization $\nu_e$
burst---a robust feature found in all modern SN simulations---can be
sufficient to fix both the neutrino hierarchy and to 
decide whether $\theta_{13}$ is ``small'' or ``large.''
\flushright{IFIC/07-01}
\end{abstract}

\maketitle

\thispagestyle{fancy}


\section{Introduction}

Despite the enormous progress of neutrino physics in the last decade, 
many open questions remain to be solved. Among them are two, the 
mass hierarchy---normal versus inverted mass spectrum---and the
value of the 13-mixing angle $\theta_{13}$, where the observation of
neutrinos from a  core-collapse supernova (SN) could provide
important clues~\cite{Dighe:1999bi,Lunardini:2003eh,Takahashi:2003bj}. 
Schematically, the neutrino emission by a SN can be divided
into four stages: Infall phase, neutronization burst, 
accretion, 
and Kelvin-Helmholtz cooling phase. 
During the infall phase and neutronization burst only $\nu_e$'s
are emitted, while the bulk of neutrino emission is released in all flavors
in the two latest phases.
Whereas the neutrino emission characteristics of the two initial
stages are basically independent of the features of the progenitor,
such as its mass or equation of state (EoS) of the core, the details of
the neutrino fluxes during the accretion and cooling
phases may  significantly change for different SN models.
Therefore, a straightforward extraction of oscillation parameters
from the bulk of the SN neutrino signal 
seems hopeless. Only features in the
detected neutrino spectra that are independent of unknown SN
parameters should be used in such an analysis.

In this talk we discuss the potential of the 
 two most promising sources for such features: the  
modulations in the $\bar\nu_e$ spectra caused by the Earth
 matter~\cite{Lunardini:2001pb,Dighe:2003jg,Dighe:2003vm} or by 
the passage of shock waves through the SN
 envelope~\cite{Schirato:2002tg,Fogli:2003dw,Tomas:2004gr}.
Moreover, we show how the detection of the
neutronization $\nu_e$ burst could help to break possible degeneracies
in the case that $\theta_{13}$ is still unknown. In
Tab.~\ref{abc-table} it is shown the dependence of these three observables
on three different neutrino mass schemes.
\begin{table}[ht]
\begin{center}
\caption{The presence of Earth-matter and shock wave effects in the
  $\bar\nu_e$ spectra and of the $\nu_e$ burst for
  different neutrino mixing scenarios.} 
\begin{tabular}{llcccc}
\hline
Case & Hierarchy &  $\sin^2 \theta_{13}$ & Earth & Shock & $\nu_e$ burst  \\
\hline
A &  Normal & $\gtrsim 10^{-3}$ & Yes & No & No \\
B & Inverted &  $\gtrsim 10^{-3}$ & No & Yes & Yes  \\
C & Any & $\lesssim 10^{-5}$ & Yes & No & Yes \\
\hline
\end{tabular}
\label{abc-table}
\end{center}
\end{table}

\section{Identifying signatures of the SN shock wave propagation}

The neutrino spectra $F_{\nu_i}$ arriving at the Earth are 
determined by the primary neutrino spectra $F^0_{\nu_i}$ as well as
the neutrino mixing scenario, 
\begin{equation}
F_{\nu_i}(E,t) = \sum_{j} p_{ji}(E,t) F^0_{\nu_j}(E,t)\, ,
\end{equation}
where $p_{ji}$ is the conversion  probability of a $\nu_j$ into $\nu_i$
after propagation through the SN mantle.
The probabilities $p_{ji}$ are basically determined by the
number of resonances that the neutrinos traverse and their
adiabaticity~\footnote{In this talk we shall not consider the possible
  presence of collective flavor neutrino conversions driven by
  neutrino-neutrino
  interactions~\cite{Duan:2006an,Duan:2006jv,Hannestad:2006nj}.}. Both
are  
directly connected to the neutrino mixing scheme. 
In contrast to the solar case, SN neutrinos must pass through
two resonance layers: the H-resonance layer at 
$\rho_{\rm H}\sim 10^3$~g/cm$^3$ corresponding to $\Delta m^2_{\rm atm}$,
and the L-resonance layer at 
$\rho_{\rm L}\sim 10$~g/cm$^3$ corresponding to $\Delta m^2_{\odot}$.
Whereas the L-resonance is always adiabatic and 
in the neutrino channel, the adiabaticity of
the H-resonance depends on the value of $\theta_{13}$, and 
the resonance shows up in the neutrino or antineutrino channel 
for a normal or inverted mass hierarchy respectively~\cite{Dighe:1999bi}.

\begin{figure}[ht]
\centering
\includegraphics[width=6.5cm,height=5.5cm]{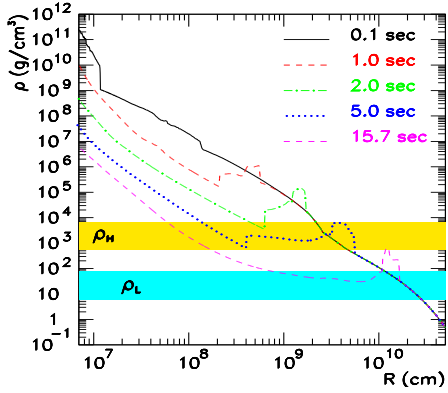}
\caption{
Density profile at several instances after core bounce.
 The resonance layers $\rho_{\rm H}$ and  $\rho_{\rm L}$ are also
 shown~\cite{Tomas:2004gr}.  
\label{fig:density}
}
\end{figure}

\begin{figure}[h!]
\centering
\includegraphics[width=6.5cm,height=5.5cm]{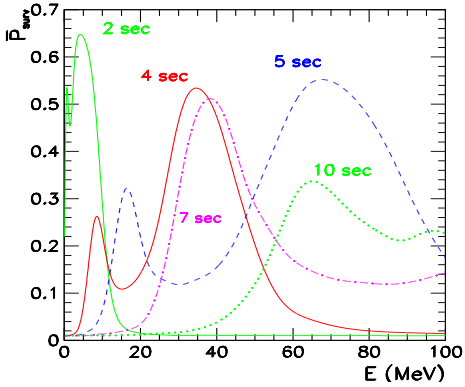}
\caption{
Antineutrino survival probability $p_{\bar\nu_e\bar\nu_e}(E,t)$ as
function of energy at different times averaging in energies with the
energy-resolution of Super-Kamiokande. A density profile with a
forward and reverse shock have been used, from
Ref.~\cite{Tomas:2004gr}.
\label{fig:pee}
}
\end{figure}

During approximately the first two seconds after core bounce, the
neutrino survival probabilities are constant in time and in energy for
all three cases A, B, and C.
However, at $t\approx 2$ s the H-resonance layer is reached by the
outgoing shock wave, see  Fig.~\ref{fig:density}. 
The way the shock wave passage affects the neutrino propagation
strongly depends on the neutrino mixing scenario: 
cases A and C will not show any evidence of shock wave propagation
in the observed $\bar\nu_e$ spectrum, either because there is no
resonance in the antineutrino channel as in scenario A, or because the
resonance is always strongly non-adiabatic as in scenario C.
However, in scenario B, the sudden change in density breaks  the
adiabaticity of the resonance, what leads to a time and energy
dependence of the antineutrino survival probability  $
p_{\bar\nu_e\bar\nu_e}(E,t)$, see Fig.\ref{fig:pee}. 
The presence of the shocks results in the appearance of bumps, one in
the case of a unique forward shock or two if an additional reverse
shock is present, in $
p_{\bar\nu_e\bar\nu_e}(E,t)$ at 
those energies for which the resonance region is passed by the shock
waves. All these structures move in time towards higher energies, as
the shock waves reach regions with lower density,
leading to observable consequences in the
$\bar\nu_e$ spectrum.

\begin{figure}[ht]
\centering
\includegraphics[width=6.5cm,height=5.5cm]{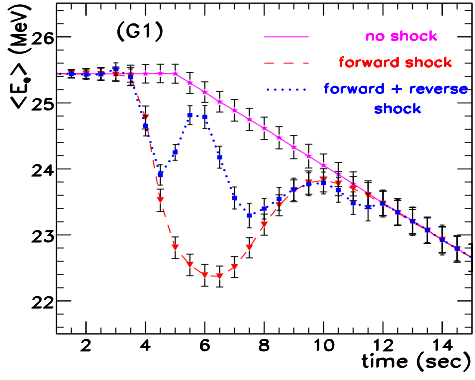}
\caption{
The average energy of $\bar\nu p\to ne^+$ events binned in time 
for a static density profile, a profile with only a forward
shock, and with forward and reverse shock. G1 model was assumed for
the neutrino fluxes. 
The error bars represent 1~$\sigma$ errors in any bin, from
Ref.~\cite{Tomas:2004gr}.
\label{fig:eav_g1}
}
\end{figure}

\begin{figure}[ht]
\centering
\includegraphics[width=6.5cm,height=5.5cm]{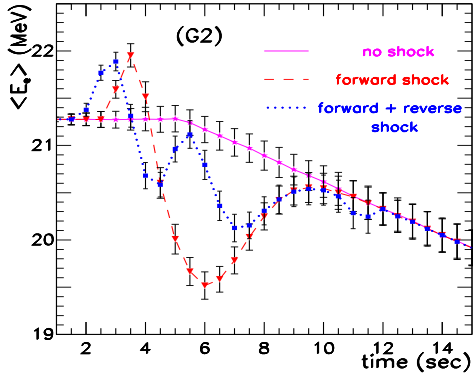}
\caption{
Same as Fig.~\ref{fig:eav_g1} for the G2 model. From 
Ref.~\cite{Tomas:2004gr}.
\label{fig:eav_g2}
}
\end{figure}

A useful observable to detect effects of the shock propagation is the
average of the measured positron energies, $\langle E_e\rangle$, produced in
inverse beta decays $\bar\nu_e+p\to n+e^+$, the most important
neutrino signal expected in a SN.
In Figs.~\ref{fig:eav_g1} and \ref{fig:eav_g2}, we show $\langle E_e \rangle$
together with the one sigma errors expected for a megaton water
Cherenkov detector and a SN in 10~kpc distance, with a time binning of
0.5~s. 
For the neutrino fluxes we use the parametrization suggested in
Refs.~\cite{Keil:2002in,Keil:2003sw},
\begin{eqnarray}
 F^0_{\nu_j}(E) & = &\frac{\Phi_0(\nu_j)}{\langle E_0(\nu_j)
 \rangle}
\frac{(\alpha_j+1)^{\alpha_j+1}}{\Gamma(\alpha_j+1)}
\left(\frac{E}{\langle
 E_0(\nu_j)\rangle}\right)^{\alpha_j} \nonumber \\
& & \exp\left(-(\alpha_j+1) \frac{E}{\langle
 E_0(\nu_j)\rangle}\right)~.
\end{eqnarray}
The values of the parameters used in the two models, $G1$ in
Fig.~\ref{fig:eav_g1} and $G2$ in Fig.~\ref{fig:eav_g2},
are given in Tab.~\ref{models-table}. The averaged energy is assumed
to decrease linearly after $5$~s.

\begin{table}[ht]
\begin{center}
\caption{The parameters of the used primary neutrino spectra models
  motivated from the SN simulations of the Garching (G1,G2) and the
  Livermore (L) group. The averaged energy is given in MeV and the
  pinching parameters $\alpha_{\bar\nu_e}$ and $\alpha_{\bar\nu_x}$
  take the values 4 and   3, respectively.} 
\begin{tabular}{cccc}
\hline
Model & $\langle E_0(\bar\nu_e) \rangle$ & $\langle E_0(\nu_x)
\rangle$ & $\Phi_0(\bar\nu_e)/\Phi_0(\nu_x)$ \\
\hline
L & 15 & 24 & 1.6 \\
G1 & 15 & 18 & 0.8 \\
G2 & 15 & 15 & 0.5 \\
\hline
\end{tabular}
\label{models-table}
\end{center}
\end{table}

The effects of the shock wave propagation are clearly visible, and are
independent of the assumptions about the initial neutrino 
spectra~\cite{Tomas:2004gr}. Moreover, it is not only possible to
detect the shock wave 
propagation in general, but also to identify the specific 
imprints of the forward and reverse shock versus the forward shock only  
case. The signature of the reverse shock is its double-dip structure
compared to the one-dip of a forward shock only.
The observation of the details of the shock wave structure will be, though,
tightly related to the absence of strong density fluctuations after
the shock front~\cite{Fogli:2006xy,Friedland:2006ta}.

\section{Earth-matter effects}

As it has been previously mentioned,
 before the shock wave reaches the H-resonance layer
 the dependence of the conversion 
probability in the cases A, B and C on the neutrino energy $E$
 is very weak. 
However, if neutrinos cross the Earth before reaching the detector, 
the conversion probabilities may become energy-dependent and induce
modulations in the neutrino energy spectrum. These
modulations  may be observed in the form of 
local peaks and valleys in the spectrum of the event rate $\sigma
F_{\bar\nu_e}^D$ plotted as a function of $1/E$. 
These modulations arise in the antineutrino channel only in cases A
and C. Therefore its observation would exclude case B.
This distortion in the spectra could be  measured by
comparing the 
neutrino signal at two or more different detectors such that the
neutrinos travel different distances through the Earth before reaching
them~\cite{Lunardini:2001pb,Dighe:2003be}. 
However these Earth matter effects can be also identified in a single
detector~\cite{Dighe:2003jg,Dighe:2003vm}.

The net $\bar\nu_e$ flux at the detector may be written in the form
\begin{eqnarray}
F_{\bar\nu_e}^D &= & \sin^2 \theta_{12}  F_{\bar\nu_x}^0 + 
\cos^2 \theta_{12} F_{\bar\nu_e}^0 \nonumber \\
& +& \Delta F^0
\sum_{i=1}^7 \bar{A}_i \sin^2(k_i y /2) \,,
\label{feDbar-y}
\end{eqnarray}
where $y$ is the ``inverse energy'' parameter $y\equiv 12.5~{\rm MeV}/E$,  
$\Delta F^0 \equiv (F_{\bar\nu_e}^0 - F_{\bar\nu_x^0})$, 
 and $\bar{A}_i$ 
depend only on the mixing parameters.

\begin{figure}[h!]
\centering
\includegraphics[width=8.5cm,height=7.5cm]{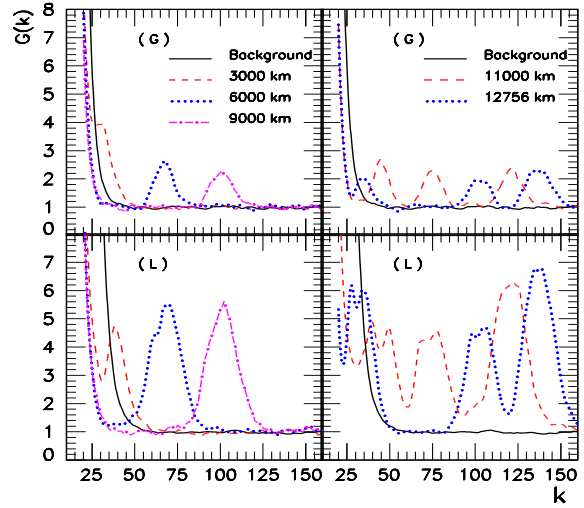}
\caption{
Averaged power spectra in the case of a large scintillator detector
for different SN models, G1 (top panels) and L (bottom panels), and
distances travelled through the Earth, from
Ref.~\cite{Dighe:2003vm}. 
\label{fig:lena_GL}}
\end{figure}

The last term in Eq.~(\ref{feDbar-y}) is the Earth oscillation term 
that contains up to seven analytically known frequencies $k_i$ in
$y$, the coefficients $\Delta F^0 \bar{A}_i$ being relatively slowly
varying functions of $y$.  The first two terms in Eq.~(\ref{feDbar-y})
are also slowly varying functions of $y$, and hence contain
frequencies in $y$ that are much smaller than the $k_i$.  
The frequencies $k_i$ are completely independent of the primary
neutrino spectra, and can be determined to a good accuracy from
the knowledge of the solar oscillation parameters, the Earth matter 
density, and the position of the SN in the sky~\cite{Dighe:2003vm}.
The latter can be determined with sufficient precision 
even if the SN is optically obscured using the pointing capability of
water Cherenkov neutrino detectors~\cite{Tomas:2003xn}.
The power spectrum of $N$ detected neutrino events is
\begin{equation}
G(k) \equiv \frac{1}{N} \left| \sum_{i=1}^N e^{iky_i} \right|^2 \,.
\label{ft-def}
\end{equation}
In the absence of Earth effect modulations, $G(k)$ has an average
value of one for $k \gtrsim 40$.  The region $k \lesssim 40$ is dominated by the
``0-peak'', which is a manifestation of the low
frequency terms in Eq.~(\ref{feDbar-y}).  Identifying Earth effects is
equivalent to observing excess power in $G(k)$ around the known
frequencies $k_i$.
In Fig.~\ref{fig:lena_GL} we show the averaged power spectra in the
case of a 32 kton scintillator detector for neutrinos travelling
only through the Earth mantle (left panels) and traversing both mantle
and core (right panels). It is possible to observe how in the former
case only one peak is present, whereas in the case that neutrinos go
through the core more frequencies arise.
The model independence of the peak positions may be confirmed by
comparing the top and bottom panels.

\begin{figure}[h!]
\centering
\includegraphics[width=6.5cm,height=5.5cm]{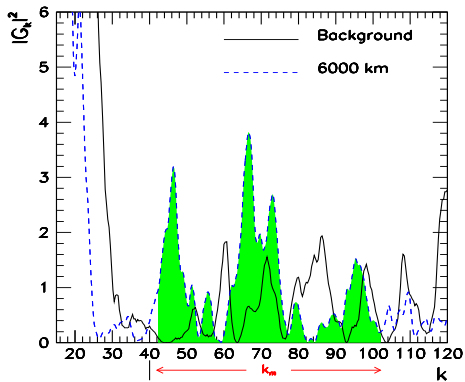}
\caption{
Realistic power spectrum from a single simulation, from
Ref.~\cite{Dighe:2003vm} . 
\label{fig:gk}}
\end{figure}

Since in the real world the presence of fluctuations in the signal,
see Fig.~\ref{fig:gk}, 
will spoil any naive theoretical peak, we need to introduce a
prescription to carry out the analysis. 
One possibility is to consider the area around the expected position
of the peak.
The area under the power
spectrum between two fixed frequencies $k_{\rm min}$ and $k_{\rm max}$
is on an average $(k_{\rm max}-k_{\rm min})$. In the absence of Earth
effects, this area will have a distribution centered around this mean.
The Earth effect peaks tend to increase this area. 
The confidence level of peak identification, $p_\alpha$, may then be
defined as the fraction of the area of the background distribution
that is less than the actual area measured.

\begin{figure}
\centering
\includegraphics[width=6.5cm,height=5.5cm]{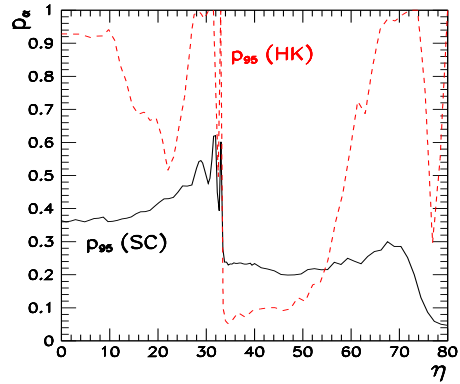}
\caption{
Comparison of $p_{95}$ as a function of nadir angle $\eta$ 
for a 32 kton scintillator (SC) and a megaton
water Cherenkov (HK) detector~\cite{Dighe:2003vm}.
\label{fig:p95}}
\end{figure}

In Fig.~\ref{fig:p95}, we compare  $p_{95}$ 
obtained with a 32 kton 
scintillator detector and a megaton water Cherenkov detector assuming
a SN at 10 kpc and the $G1$ model for neutrino fluxes.
  In the 
latter case, as neutrinos travel more and more distance in the mantle
the peak moves to higher $k$ values, and due to the high $k$
suppression, the efficiency of
peak identification decreases. When the neutrinos start
traversing the core, additional low $k$ peaks are generated and the
efficiency increases again.
The identification of Earth matter effects excludes case B, and is
thus complementary to the observation of shock wave effects.

\section{Neutronization ${\bf \nu_e}$ burst}

If the value of $\theta_{13}$ is unknown, a degeneracy exists between
case A and C. Both scenarios predict the same $\bar\nu_e$ signature
in a water Cherenkov detector, and therefore the previous two
observables are not useful to disentangle them.
In this case, the additional information
encoded in the $\nu_e$ neutrinos emitted during the neutronization
burst can fix the range of $\theta_{13}$ as well as the neutrino
mass hierarchy.  

\begin{figure}[h]
\centering
\includegraphics[width=6.5cm,height=5.5cm]{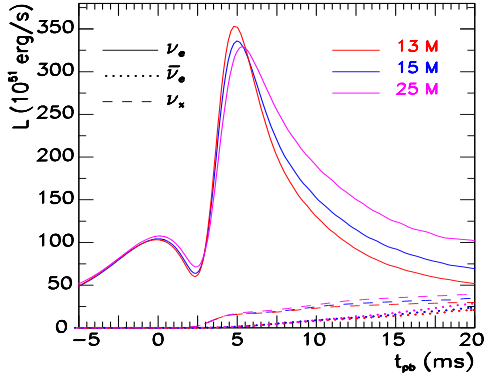}
\caption{
Neutrino luminosities as function of time for different
progenitor masses, from Ref.~\cite{Kachelriess:2004ds}.
\label{fig:nuepeak}}
\end{figure}

The prompt neutronization burst takes place during the first $\sim$
25~ms after the core bounce with a typical full width half maximum of
5--7$\,$ms and a peak luminosity 
of 3.3--3.5$\times 10^{53}\,$erg$\,$s$^{-1}$. The striking
similarity of the neutrino emission characteristics 
despite the variability in the properties of the
pre-collapse cores is caused by a 
regulation mechanism between electron number fraction and
target abundances for electron capture which establishes similar
electron fractions in the inner core 
during collapse. 
This leads to a convergence of the structure
of the central part of the
collapsing cores and only small differences in the evolution of
different progenitors until shock breakout.
The small dependence of the neutronization burst on, e.g., the progenitor
mass can be verified in Fig.~\ref{fig:nuepeak} (cf. also
Refs.~\cite{Kachelriess:2004ds,Takahashi:2003rn}).

Taking into account that the SN will be likely obscured by dust and a good 
estimation of the distance will not be possible, 
the time structure of the detected neutrino signal should be
used as signature for the neutronization burst~\cite{Kachelriess:2004ds}. 
Since the event number in current and proposed charged-current $\nu_e$ 
detectors is not high enough to allow for a detailed time analysis, 
we discuss only the case of a megaton water Cherenkov detector.
Here
one has to consider the $\nu_e$ elastic scattering on
electrons, which is affected by
several backgrounds like inverse beta decay or reactions on
oxygen. This background can
be substantially reduced by using angular and energy cuts, as well as
Gadolinium to tag neutrons from inverse beta decays. 
The sample of elastic scattering events still contains the irreducible
background of scattering on electrons of other neutrino flavors than
$\nu_e$, but this contamination does not affect the possibility to
disentangle the different neutrino scenarios~\cite{Kachelriess:2004ds}.

The time evolution of the signal depends strongly on the neutrino
mixing scheme. In case A, the $\nu_e$ survival probability  is
close to zero, and therefore the peak structure observed in the initial
$\nu_e$ luminosity is absent. On the contrary, in case C, 30\% of the
original $\nu_e$ remain as $\nu_e$  whereas 70\%
are converted into $\nu_x$. Since the cross section of $\nu_e$ on
electrons is much larger than that of $\nu_x$, the signal is dominated
by the contribution of $\nu_e$. These $\nu_e$'s follow the time evolution of
$L_{\nu_e}$, and thus lead to a clear peak in the signal.

\begin{figure}[h!]
\centering
\includegraphics[width=6.5cm,height=5.5cm]{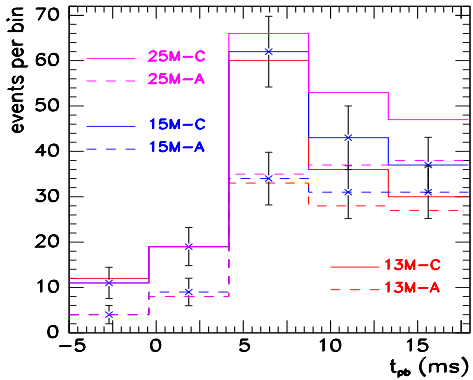}
\caption{
 Number of $\nu e\rightarrow \nu e$ events per time bin  in a megaton
water Cherenkov detector for a SN at 10 kpc for cases A (dashed lines)
and C (solid lines) and for different progenitor masses. Statistical
errors are also shown for the $15~M_\odot$ case~\cite{Kachelriess:2004ds}. 
\label{fig:nuepeak2}}
\end{figure}
In Fig.~\ref{fig:nuepeak2} we show the expected neutrino
signal from $t=-5$ to 18~ms for different progenitor masses, and for
the mixing scenarios A and C. The peak structure can be clearly seen
in case C, but not in case A~\cite{Kachelriess:2004ds}.  Including
recent improvements of the electron capture rates or uncertainties in
the nuclear equation of state has only little effect on the
neutronization peak  compared to the size of the
statistical fluctuations. 
Therefore the observation of a peak in the
first milliseconds of the neutrino signal would rule out case A.

\section{Summary}

A reliable determination of neutrino parameters using SN neutrinos
should be independent from the primary neutrino fluxes produced during
the accretion and cooling phase of the SN. 
Earth-matter effects and the passage of SN shocks through the
H-resonance both introduce unique modulations in the neutrino energy
spectrum that allow one their identification without knowledge of the
primary neutrino spectra.
While the observation of Earth-matter effects in the $\bar\nu_e$
energy spectrum rules out case B, modulations in the $\bar\nu_e$ time
spectrum identify  case B. If the value of $\theta_{13}$ would be
known to be large, then the neutrino mass hierarchy would be
identified. Otherwise, the detection of the neutronization $\nu_e$
peak---a robust feature of all modern SN simulations---can break the
remaining degeneracy between A and C.

\bigskip 
\begin{acknowledgments}
I am greatly grateful to the local organizers for the warm hospitality during
the Conference.
I would also like to thank M.~Frigerio, G.~Senjanovic, and S.~Thomas for very
pleasant and interesting discussions.
The results presented in this talk were derived in collaboration with 
M.~Kachelrie\ss, A.~S.~Dighe,
G.~G.~Raffelt, H.~-Th. Janka, L.~Scheck, R.~Buras, A.~Marek, and
M.~Rampp.
This work has been supported by the Juan de la Cierva  programme,
an ERG from the European Comission, and by the Spanish grant FPA2005-01269.
\end{acknowledgments}

\bigskip 

\end{document}